# Crack Tip Relaxation Governs Onset of Fracture Instability

Farid F. Abraham

Lawrence Livermore National Laboratory

Livermore, CA 94550

**Abstract:**

A dynamic crack will travel in a straight path up to a material-dependent critical speed beyond which its path becomes erratic. Predicting this critical speed and discovering the origin of this instability are two outstanding problems in fracture mechanics. We recently discovered a simple scaling model based on an *effective* elastic modulus that gives successful predictions for this critical speed by transforming the nonlinear crack dynamics problem into a linear elasticity representation. We now show that a simple atomic picture based on broken-bond relaxation at the dynamic crack tip provides an explanation for the origin of the effective elastic modulus.

In 1951, Yoffe (*1*) made the physically intuitive suggestion that mode I crack growth occurs in the direction of maximum asymptotic hoop stress and found the crack speed for the onset for branching to be about 70% of the Rayleigh wave speed $c_R$ (*2, 3*). However, this high speed is rarely observed in experiment (*4, 5*). An obvious shortcoming in Yoffe's analysis is the assumption of a constant linear elastic response for all deformations. In our recent study of brittle fracture (6), we showed that hyperelasticity, the elasticity at large strain, plays a governing role in the onset of the crack instability from unidirectional motion. We discovered a simple, yet remarkable, scaling based on an *effective* elastic modulus for our modelled solid (the secant modulus at the stability limit of the bulk solid), which led to successful predictions for the onset speed of the crack instability. We have also applied this scaling to the same-modelled solid with the exception that the crack is constrained to travel unidirectional irrespective of its speed (7). This allowed the crack to achieve a unique steady-state speed that has a dependence on hyperelasticity. Using our scaling law, we found that the steady-state crack speed scales to a constant value equal to a crack speed of a linear solid with our *effective* elastic modulus. In this paper, we demonstrate that atomic relaxation of breaking bonds at the crack tip governs these dynamic features of the travelling brittle crack.

We summarize our earlier findings. Our simulation model is based on a generalized bilinear force law composed of two spring constants, one associated with small deformations ($k_1$, $r < r_{on}$) and the other associated with large deformations ($k_2$, $r > r_{on}$). This is shown in Figure 1(a). This model allowed us to investigate the generic effects of hyperelasticity by changing the relative magnitude of the spring constants $\alpha = k_2/k_1$ and transition distance $r_{on}$ of the potential [in terms of $\varepsilon_0 = (r_{on}/r_0) - 1$]. We considered the propagation of a crack in two-dimensional hexagonal lattice geometry. The slab is loaded in mode I with a constant strain rate. The dynamic crack instabilities for the various $\alpha = k_2/k_1$ are associated with the precipitous drops in crack speed (see Figure 1b), as indicated by the arrows, and are a consequence of the crack deviating from straight line motion (see Figure 1c). The crack speed at the onset of erratic motion is defined as the instability speed.

Figure 2a presents a log-log plot of the instability speed as a function of $\alpha = k_2/k_1$ for various $\varepsilon_0 = (r_{on}/r_0)-1$. For each $\varepsilon_0$, we found that the dependence is essentially linear, the slope approaching one-half for $\varepsilon_0$ tending to zero. This trend is required since $k_2/k_1 = 1$ for $\varepsilon_0 = 0$ and the solid is strictly linear with a spring constant equal to $k_2$. Therefore, the instability speed will have a trivial square-root dependence on the spring constant $k_2$ when normalized by $k_1$. The other limit is $r_{on} = r_{break}$. In this limit, the bilinear force law is simply the linear force with spring constant $k_1$. Figure 2b defines, graphically, our choice for an *effective* spring constant $k_{eff}$ of the bilinear force law. The elastic modulus associated with this effective spring constant is the *secant modulus* at the mechanical stability limit. By plotting in Figure 2c the instability speed as a function of $\alpha_{eff} = k_{eff}/k_1$, we see a remarkable collapse of the data from Figure 2a onto a common straight line with slope equal to one-half. For determining the instability speed of a dynamic brittle crack, this finding allows one to model the bilinear material as a linear solid with the effective spring constant just described. We applied this concept of an effective spring constant to a continuous interatomic potential: in particular, to the Lennard-Jones 12:6 potential. The prediction is in agreement with computer simulations (8, *9*).

For a simple linear solid, the instability speed is 0.73 in agreement the Yoffe prediction. For a nonlinear solid, the instability speed is $0.73(k_{eff} / k_1)^{1/2}$. This suggests that Yoffe's picture of the dynamic instability in brittle fracture may be valid. It is only necessary to replace the elastic modulus for small deformation with an *effective* elastic modulus (the secant modulus) described in this study, giving successful predictions for the onset speed of the crack instability for nonlinear materials.

Abraham et al. (*8, 9*) proposed that the onset of the instability can be understood from the point of view of reduced local lattice vibration frequencies due to softening at the crack tip. They noted that the onset of the roughening (the instability) corresponds to the point in the crack tip dynamics where the time it takes the tip to transverse one lattice spacing approximately equals the period of one atomic vibration. Hence, they propose that the bond-breaking process no longer "sees" a symmetric environment due to thermal averaging, but begins to experience local atomic configurations "instantaneously" distorted from the perfect lattice symmetry. They suggested that this gives rise to small scale atomic fluctuations in the bond-breaking path and, hence, atomic roughening. This symmetry breaking results in atomic roughening of the crack path and triggers larger scale deviations with growing crack length.

Marder (*10*) discovered the importance of the atomic vibrations at the crack tip in explaining the phenomenon of lattice trapping and the velocity gap associated with the initiation of crack motion. We will quote his discussion since it lends important insights into crack dynamics that demand incorporating atomic scale behavior. *"Dynamic fracture is a cascade of bonds breaking, one giving way after another like a toppling line of dominos.* Figure 3a *shows what happens as the crack moves forward. In the second frame, the bond between two atoms has just broken. There is no guarantee that the next bond to the right will break. The crack could fall into a static lattice-trap state. The best chance to avoid this fate is for the atom marked in green to deliver enough of a blow to its right-hand neighbor that the bond on that neighbor also breaks. This process must take place within the first half of the first* *vibrating period of the green atom. This is because the longer it vibrates, the more energy is dispersed to its neighbors in all directions in the form of traveling waves. This dispersal decreases the chance that there will be enough concentrated energy available to snap the next bond down the line."* Because of this upper limit on the time interval between breaking consecutive bonds, one should expect a lower limit on the propagation speed of rapid cracks.

Both discussions emphasize the importance of considering the atomic dynamics at the crack tip. Our original picture for the origin of the crack instability and Marder's study suggest that the following important question should be addressed: "How fast does a 'snapping bond' at the crack tip relax?" A sensible approximation for answering this question is to measure the time it takes for a single atom coupled to a spring, obeying our bilinear force law, to move from a fully extended state, $r_{break}$, to the unstretched state, $r_0$. We determined this time numerically for a variety of $a$ and $a_0$ combinations and express it as a measured spring constant $k_{bilinear}$. In Figure 3b, we note excellent correlation with the $k_{eff}$, clearly showing that a measure of the relaxation of the single atom driven by the bilinear spring is well approximated by a linear spring with spring constant $k_{eff}$. This finding, along with the picture that the atomic relaxation (vibration) is the origin of the dynamic instability, is consistent with explaining the "remarkable scaling" shown in Figure 2c.

This effective spring constant may be interpreted as specifying an *effective wave velocity* $c_{eff}$ for energy transfer between breaking bonds at the crack tip. We know that Yoffe's solution gives the correct instability speed for a linear solid. Identifying the linear wave speed in Yoffe's solution with $c_{eff}$ gives a generalization of Yoffe's theory where account for hyperelasticity is included.

In summary, a coherent physical picture describing the origin of dynamic crack roughening in brittle fracture has evolved. The hyperelasticity, or elasticity at large strain, plays a governing role in the instability dynamics. A simple scaling model based on an *effective* elastic modulus has been discovered and gives successful predictions for the onset speed of the brittle crack instability by transforming the nonlinear crack dynamics problem into a linear elasticity representation. An atomic picture based on broken-bond relaxation at the dynamic crack tip provides an understanding for the origin of the effective elastic modulus. The development of a first-principles theory remains a theoretical challenge.

**Acknowledgement**

Work was performed under the auspices of the U.S. Department of Energy by the University of California, Lawrence Livermore National Laboratory under contract W-7405-Eng-48.

**Figure Captions**

**Figure 1**. (a) The bilinear force is composed of two spring constants, one associated with small deformations ($k_1$ for $r < r_{on}$) and the other associated with large deformations ($k_2$ for $r > r_{on}$). The Lennard-Jones force law is shown as a dotted blue line. (b) A crack speed history is depicted for the bilinear solid for a particular $\alpha = k_2/k_1$ and transition distance expressed as $\varepsilon_0 = (r_{on}/r_0) - 1$. The dynamic crack instability is indicated by the arrow. (c) A picture of a crack is shown at a significant time beyond the onset of the instability.

**Figure 2.** a) A log-log plot of the instability speed as a function of $\alpha = k_2/k_1$ is presented for various $\varepsilon_0 = (r_{on}/r_0) - 1$. b) The *effective* spring constant $k_{eff}$ is defined graphically for the bilinear force. c) The instability speed is presented as a function of $\alpha_{eff} = k_{eff}/k_1$ and shows the remarkable collapse of the data to a simple square-root dependence. Application of the scaling to the continuous Lennard-Jones potential is demonstrated.

**Figure 3.** a) Dynamic fracture is a cascade of bonds breaking, one giving way after another like a toppling line of dominos. b) We note excellent correlation with the $k_{eff,}$ clearly showing that a measure of the relaxation of the single atom driven by the bilinear spring is well approximated by a linear spring with spring constant $k_{eff.}$

# Figure 1

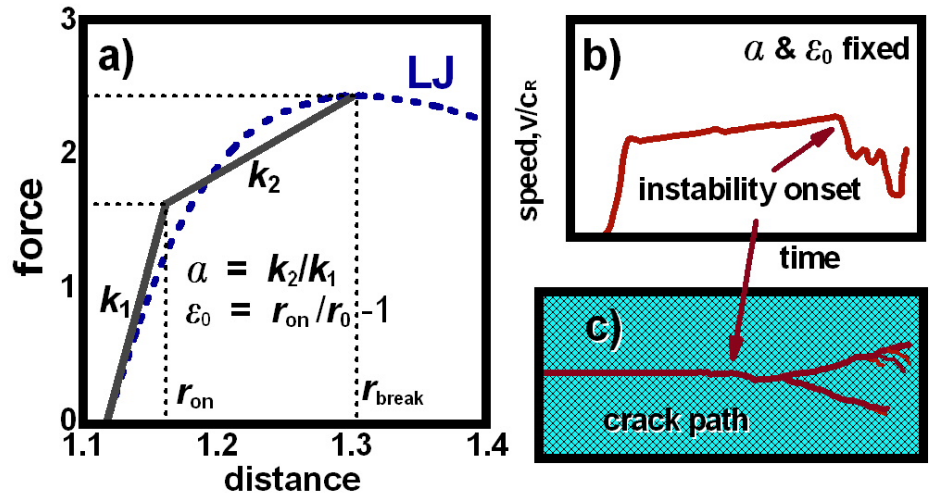

**Figure 2**

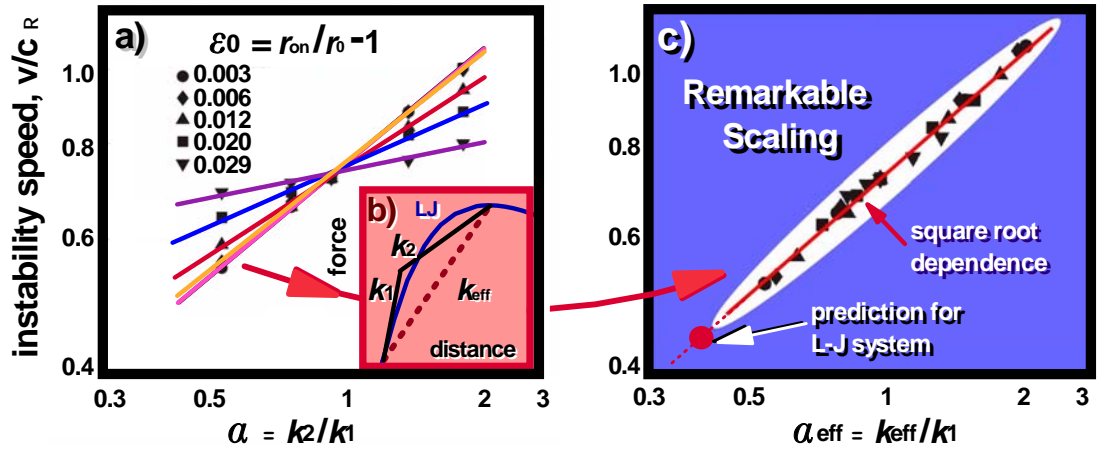

# Figure 3

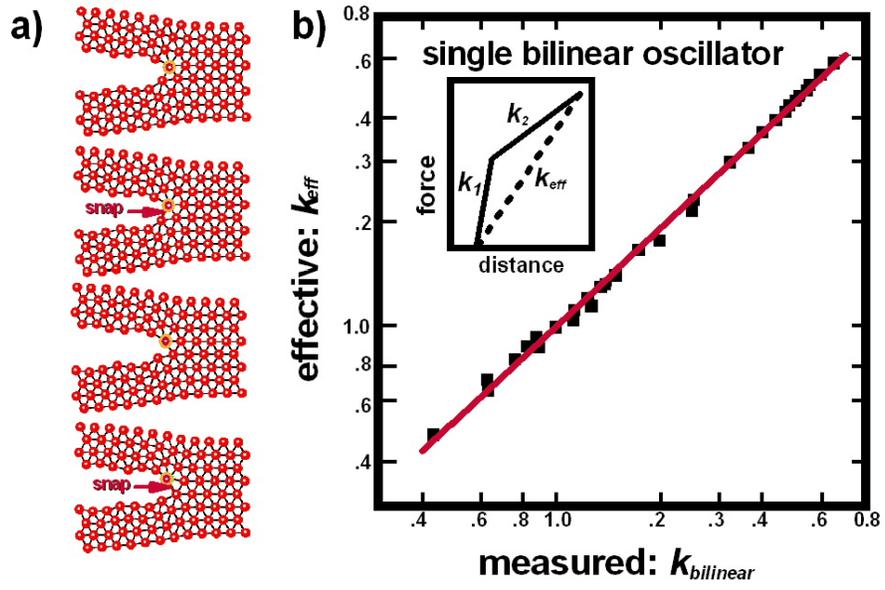